\documentclass{PoS}
\usepackage{epsfig}

\newcommand{\xText}[3]{\put(#1,#2){#3}}
\newcommand{\be}{\begin{equation}}

\newcommand{\ee}{\end{equation}}
\newcommand{\bea}{\begin{eqnarray}}
\newcommand{\eea}{\end{eqnarray}}

\newcommand{\eps}{\epsilon}

\def\eps{{\epsilon}}

\title{The BESS model revisited as a Higgsless Linear Moose @ the LHC}

\ShortTitle{The BESS model @ the LHC}

\author{\speaker{Stefania De Curtis}\thanks{In collaboration with E. Accomando, D. Dominici, and L. Fedeli}\\
        INFN, 50019 Sesto F., Firenze, Italy\\
        E-mail: \email{decurtis@fi.infn.it}}

\abstract{We study the phenomenological consequences of a four site
Higgsless model based on the $SU(2)_L\times SU(2)_1\times
SU(2)_2\times U(1)_Y$ gauge symmetry, which predicts two neutral and
four charged extra gauge bosons, $Z_{1,2}$ and $W^\pm_{1,2}$. The
model represents an extension of the minimal three site version (or
BESS model), largely  investigated in the literature, which includes
three heavy vector bosons.  We compute the properties of the new
particles, and derive indirect and direct limits on their masses and
couplings from LEP and Tevatron data and from the perturbative
unitarity requirements. In contrast to other Higgsless models
characterized by fermiophobic extra gauge bosons, here sizeable
fermion-boson couplings are allowed by the electroweak precision
data. The prospects of detecting the new predicted particles in the
favoured Drell-Yan channel at the LHC are thus investigated. The
outcome is that all six extra gauge bosons could be discovered in
the early stage of the LHC low-luminosity run.}

\FullConference{2008 Physics at LHC\\
         September 29 - 4 October 2008\\
         Split, Croatia}

\begin{document}

\section{Introduction}
During the last years a remarkable activity has been devoted to
investigate Higgsless  models  because they emerge in a natural way
when considering local gauge theories in five dimensions. Their
major outcome consists in delaying the unitarity violation of vector
boson scattering amplitudes to higher energies compared to the
answer of the Standard Model (SM) without a light Higgs, via the
exchange of Kaluza-Klein (KK) excitations. The discretization of the
compact fifth dimension to a lattice generates the so-called
deconstructed theories which are chiral lagrangians with a number of
replicas of the gauge group equal to the number of lattice sites
\cite{decon}. The drawback of all these models, as with technicolor
theories, is to reconcile the presence of a relatively low
KK-spectrum, necessary to delay the unitarity violation to
TeV-energies, with the electroweak precision tests (EWPT) whose
measurements can be expressed in terms of the $\eps_1,\eps_2$ and
$\eps_3$ (or $T, U, S$) parameters. This problem can be solved by
either delocalizing fermions along the fifth dimension
\cite{Cacciapaglia:2004rb}
 or, equivalently in the deconstructed
picture, by allowing for direct couplings between new vector bosons
and SM fermions \cite{Casalbuoni:2005rs}. In the simplest version of
this latter class of models, corresponding to just three lattice
sites and gauge symmetry $SU(2)_L\times SU(2)\times U(1)_Y$ (the
so-called BESS model \cite{Casalbuoni:1985kq}), the requirement of
vanishing of the $\eps_3$ parameter implies that the new triplet of
vector bosons is almost fermiophobic, then the only production
channels for their search are those driven by boson-boson couplings.
The Higgsless literature has been thus mostly focused on difficult
multi-particle processes which require high luminosity to be
detected, that is vector boson fusion and associated production of
new gauge bosons with SM ones. We extend the minimal three site
model by inserting an additional lattice site. This new four site
Higgsless model, based on the $SU(2)_L\times SU(2)_1\times
SU(2)_2\times U(1)_Y$ gauge symmetry, predicts two neutral and four
charged extra gauge bosons, $Z_{1,2}$ and $W^\pm_{1,2}$, and
satisfies the EWPT constraints without necessarily having
fermiophobic resonances \cite{Accomando:2008jh}. Within this
framework, the more promising Drell-Yan processes become
particularly relevant for the extra gauge boson search at the LHC.

\section{The model: Unitarity and EWPT bounds}

The class of models we are interested in, called linear {\it{moose}}
models, follows the idea of dimensional deconstruction
 \cite{ decon}.  In their general
formulation they are based on the  $SU(2)_L\otimes SU(2)^K\otimes
U(1)_Y$ gauge symmetry, and contain $K+1$ non linear $\sigma$-model
scalar fields $\Sigma_i$,  interacting with the
 gauge fields. The
new parameters are the $K$ gauge coupling constants $g_i$ and the
$K+1$ link couplings $f_i$. Different $f_i$, in the continuum limit,
can describe a generic warped metric. For details see
\cite{Casalbuoni:2004id}. Direct couplings of new gauge bosons to SM
fermions can be included in a way that preserve the symmetry of the
model. According to \cite{Casalbuoni:2005rs}, we consider only
direct couplings of the new gauge bosons to the left-handed fermions
with strength  given by the parameters $b_i$. The case $K=1$
corresponds to the BESS model \cite{Casalbuoni:1985kq}.  Here we
concentrate on the case $K=2$. For simplicity we will assume
$g_1=g_2$ and $f_1=f_3$. This choice corresponds to a L-R symmetry
in the new gauge sector, leading to a definite parity for the
corresponding gauge bosons once the standard gauge interactions are
turned off.  The charged and neutral gauge boson spectrum and their
couplings to fermions are given in \cite{Accomando:2008jh}. Summing
up, the parameters of the four site model are $g_1$, $f_1$, $f_2$,
$b_1$ and $b_2$. In order to reproduce the physical values of the SM
gauge boson masses, we get a relation among the parameters of the
model and we end with four free parameters: $b_1$, $b_2$ and $M_1$,
$M_2$ (related to $f_1$, $f_2$ and equal to, apart from weak
corrections, the mass eigenvalues of the two new triplet of gauge
bosons). We will call $z$ the ratio $M_1/M_2$, with $z<1$.

\begin{figure}[!h]
\begin{center}
\includegraphics[width=4.3 cm]{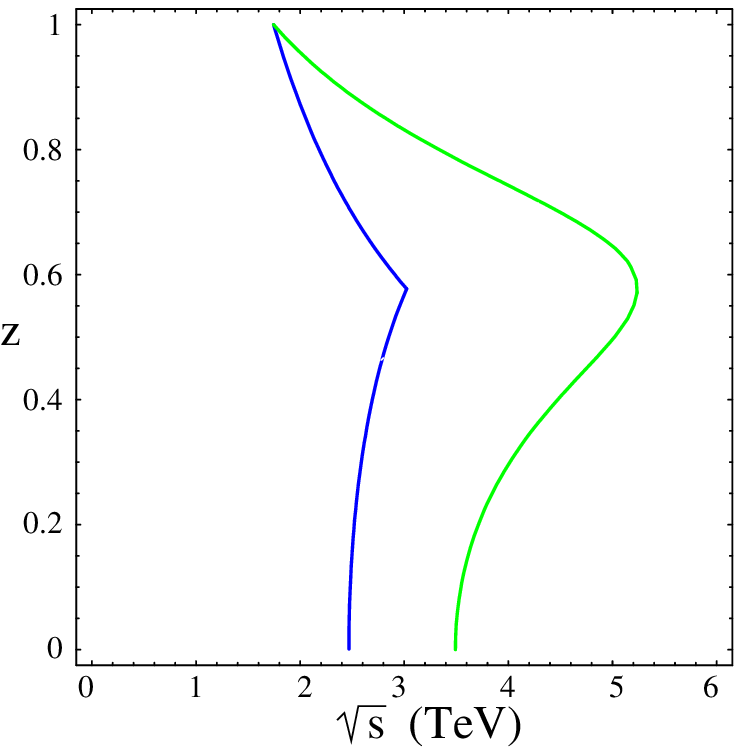}~~~~~~~~~
\includegraphics[width=4.5 cm]{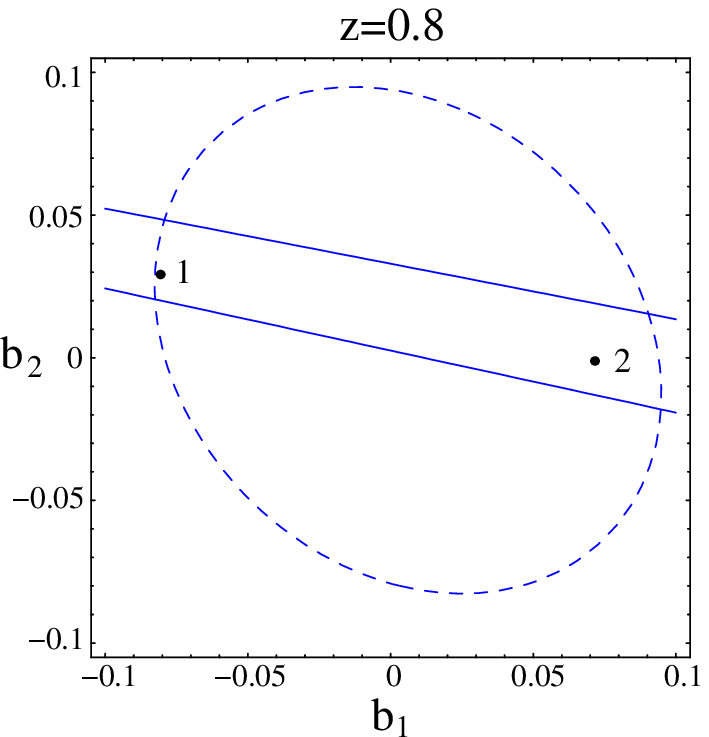}
\end{center}
\caption{Left: Unitarity bounds on the plane ($\sqrt{s}$, $z$), for
$W_L W_L$ scattering amplitudes (green lighter curve), and for
scattering amplitudes with external SM and extra gauge bosons (blue
darker curve). The allowed regions are on the left of the curves.
Right: 95\% C.L. bounds on the plane $(b_1,b_2)$ from $\epsilon_1$
(dash line) and $\epsilon_3$ (solid line) for $z=0.8$ and $500\le
M_1({\rm GeV})\le 1000$. The allowed regions are the internal ones.
The points {\bf 1} and {\bf 2} correspond to the two scenarios
considered in the following phenomenological analysis.}\label{bi1}
\end{figure}

The four site Higgsless model is an effective description.  The
energy scale where the perturbative regime is still valid is plotted
in Fig. \ref{bi1}. Owing to the exchange of the extra gauge bosons,
the unitarity violation for the $WW$ (all channels) longitudinal
scattering amplitudes can be delayed up to $\sqrt{s}=5(3)$TeV
\cite{Accomando:2008jh}. Hence, the mass spectrum of the new
particles is constrained to be within a few TeV. In general, the
only way to combine the need of relatively low mass extra gauge
bosons with EWPT is to impose the new particles to be fermiophobic.
In our four site model this assumption is not necessary anymore.  We
get bounds on the parameter space of the model by deriving the new
physics contribution to the electroweak parameters $\epsilon_1$,
$\epsilon_2$ and $\epsilon_3$ \cite{Barbieri:2004qk} and comparing
with their experimental values. The result is shown in Fig.
\ref{bi1} for  $z=0.8$ and $500\le M_1({\rm GeV})\le 1000$. Here the
bound from $\epsilon_3$ goes from up to down by increasing $M_1$,
that from $\epsilon_1$ is quite insensitive to the value of the
resonance masses ($\epsilon_2$ doesn't give any relevant limit
thanks to its negative experimental value). We see that, while the
relation between the couplings $b_1$ and $b_2$  is strongly
constrained by the $\epsilon_3$-parameter, their magnitude is weakly
limited by $\epsilon_1$. As a result, the direct fermion-boson
couplings can be of the same order of the SM ones. The
phenomenological consequence is that the four site Higgsless model
could be proved in the more promising Drell-Yan channel already at
the LHC start-up.

\section{Extra gauge boson production in Drell-Yan channels at the LHC}

\begin{figure}[t]
\begin{center}
\unitlength1.0cm
 \begin{picture}(8,6)
  \xText{-2.5}{5.3}{\small{~~~~$z=0.8$,~~~$M_{1,2}=(1000,
  1250)$GeV}}
  \xText{-1.3}{-0.1}{${M}_{inv}(l^+l^-)$[GeV]}
  \xText{-4.}{3.}{$N_{evt}$}
\xText{5.2}{5.3}{\small{~~~~$z=0.8$,~~~$M_{1,2}=(1000, 1250)$GeV}}
  \xText{6.7}{-0.1}{${M}_t(l\nu_l) $[GeV]}
  \xText{3.9}{3.}{$N_{evt}$}
\put(-4.2,-0.7){\epsfig{file=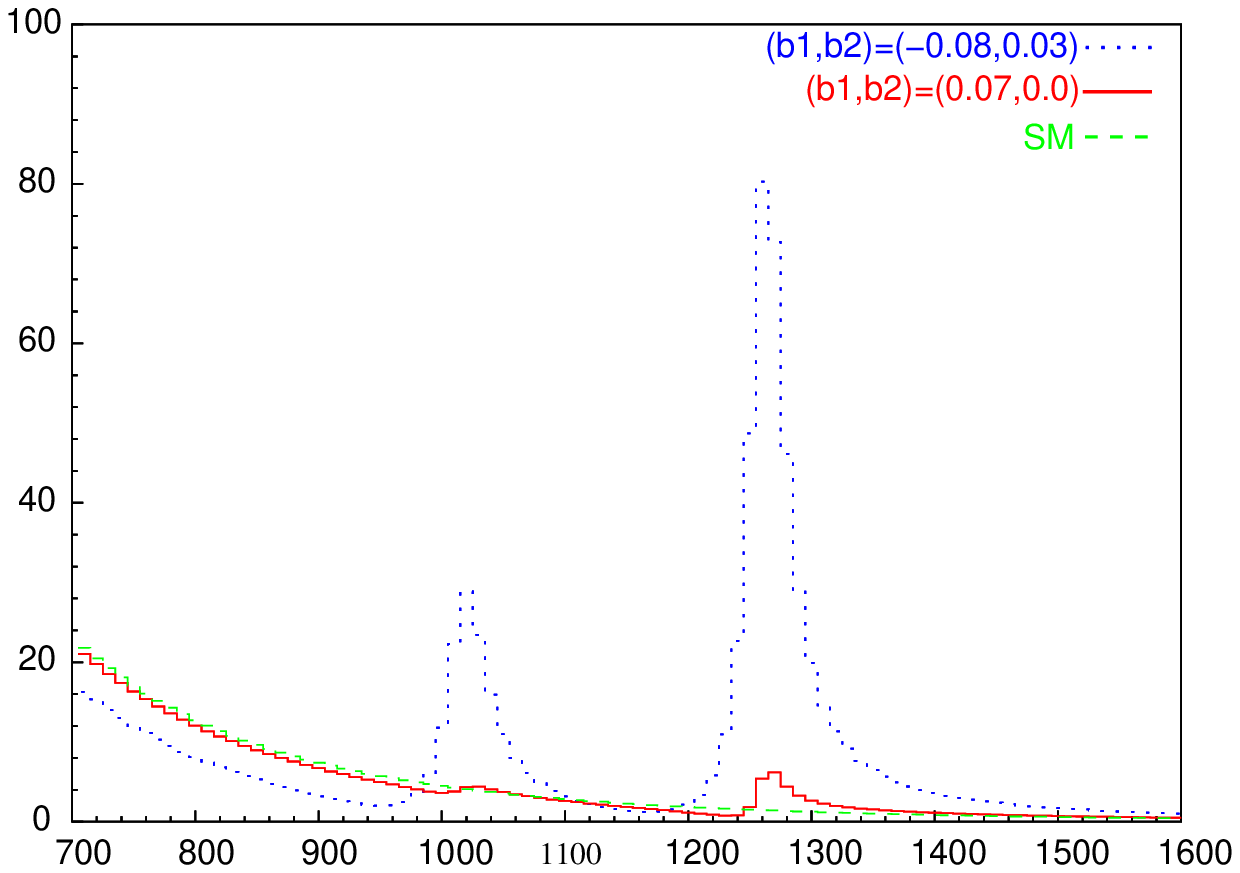,width=12.cm}}
  \put(3.6,-0.7){\epsfig{file=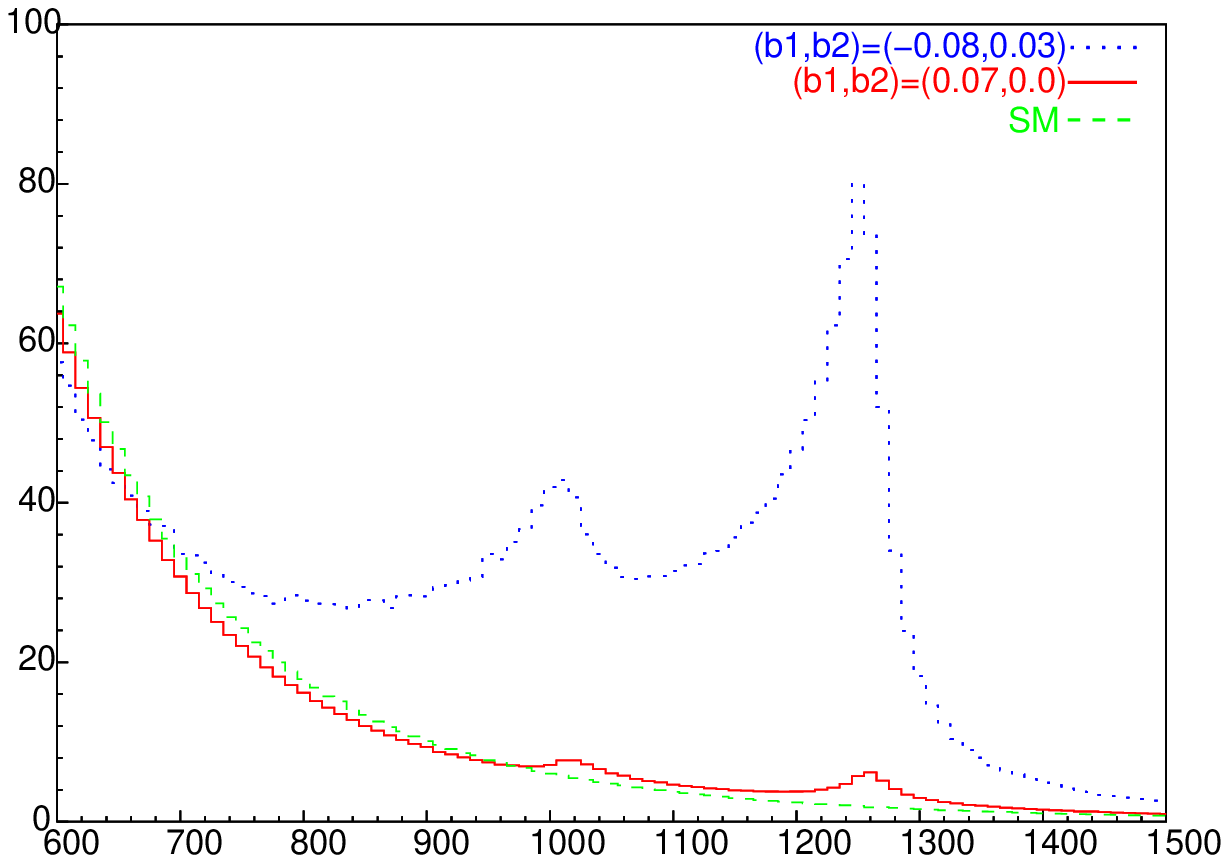,width=12.cm}}
\end{picture}
\end{center}
\vskip .1cm \caption{Total number of events in a 10 GeV-bin versus
the dilepton invariant mass, $M_{inv}(l^+l^-)$ for the process $p
p\rightarrow l^+l^-$ (left) and versus the lepton transverse mass,
$M_t(l\nu_l)$ (right), for the process $p p\rightarrow l\nu_l$ at
the integrated luminosity $L=10$ fb$^{-1}$ for $M_{1,2}=(1000,
1250)$GeV and the two scenarios {\bf 1}: $(b_1, b_2)=(-0.08,0.03)$
and {\bf 2}: $(b_1, b_2)=(0.07,0.0)$. We sum over $e,\mu$ and charge
conjugate channels. } \label{fig:minv}
\end{figure}

Let's now consider the production of the  six new gauge bosons
$Z_{1,2}$ and $W^\pm_{1,2}$ predicted by the four site Higgsless
model at the LHC through Drell-Yan channels. We analyze two classes
of processes:  $pp\to l^+l^-$ and $pp\to l\nu_l$,   with $l=e,\mu$.
The first class is characterized by two isolated charged leptons in
the final state. The latter gives instead rise to one isolated
charged lepton plus missing energy. The aforementioned neutral and
charged Drell-Yan channels can involve the production of two neutral
extra gauge bosons, $Z_1$ and $Z_2$, and four charged extra gauge
bosons $W_1^\pm$ and $W_2^\pm$ as intermediate states, respectively.

Our numerical setup is here summarized: $M_Z=91.187$GeV, $\alpha
(M_Z)=1/128.8$, $G_\mu =1.1664\cdot 10^{-5}{\rm GeV}^{-2}$. We adopt
the fixed-width scheme and apply standard acceptance cuts:
$P_t(l)>20$GeV, $P_t^{miss}>20$GeV, $\eta (l)<2.5$.  The
distribution functions CTEQ6L are used. As an example of the four
site Higgsless model prediction, we choose two sets of free
parameters: setup {\bf 1}: $(b_1, b_2)=(-0.08,0.03)$ and setup {\bf
2}: $(b_1, b_2)=(0.07,0)$ at fixed values $M_{1,2}=(1000,1250)$GeV,
corresponding to $z=0.8$. We compute the full Drell-Yan process,
considering signal and SM-background, at EW and QCD leading order. A
luminosity L=10 fb$^{-1}$ is assumed. In Fig. \ref{fig:minv}, we
plot the distributions both in the charged and neutral Drell-Yan
channels. For example, the number of signal (total) events in the
range $|M_{inv}(l^+l^-)-M_{1,2}|<\Gamma_{1,2}$ for the neutral
channel is $N_{evt}(Z_1)=108(119)$ and $N_{evt}(Z_2)=291(302)$ for
the setup {\bf 1} and $N_{evt}(Z_1)=3(28)$ and $N_{evt}(Z_2)=15(22)$
for the setup {\bf 2}. The results show that, while the  setup {\bf
2} would need high luminosity, in the first one the new gauge bosons
could be discovered already at the LHC start-up, with a minimum
luminosity L=1 fb$^{-1}$. A detailed analysis is given in
\cite{Accomando:2008jh}.
\begin{figure}[!h]
\begin{center}
\includegraphics[width=5.5 cm]{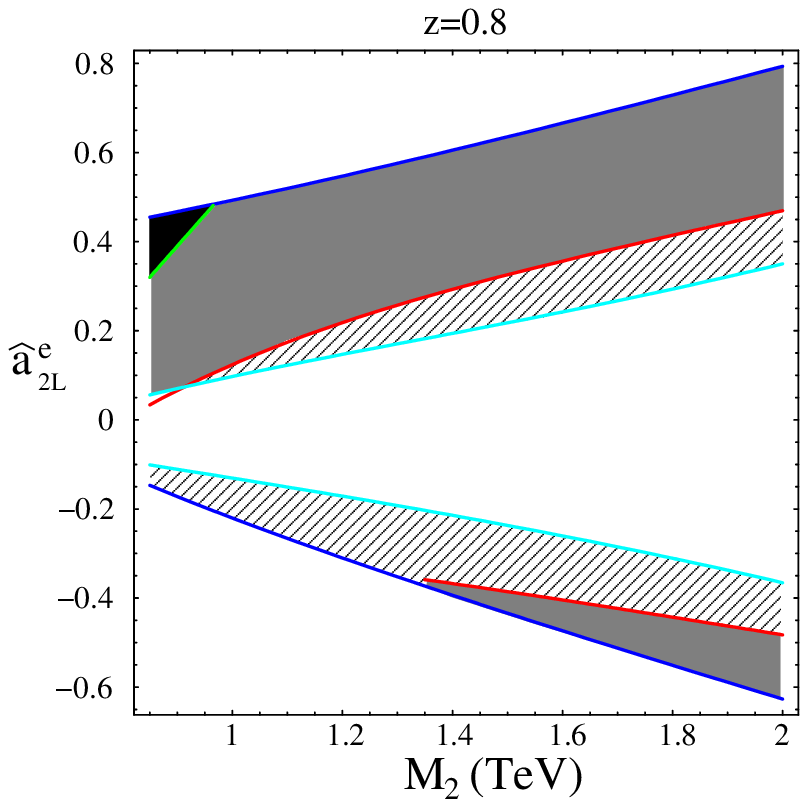}
\end{center}
\caption{5$\sigma$-discovery plot at L=100 fb$^{-1}$ in the plane
 $({\hat a}^e_{2L}, M_2)$  for $z=0.8$.
 The upper and lower parts are excluded by EWPT, the
black triangle  is the region excluded by the direct search at the
Tevatron for a luminosity of 4 fb$^{-1}$. Inside the dark-grey
regions both $Z_{1,2}$ are visible; inside the grey (dashed) ones
only $Z_1$ ($Z_2$)  can be detected. Inside the central uncolored
region no resonance is visible in the Drell-Yan channel.}
\label{fig:visibilita}
\end{figure}

In Fig.~\ref{fig:visibilita} we plot the 5$\sigma$-discovery
contours at L=100 fb$^{-1}$ in the plane $({\hat a}_{2L}^e, M_2)$,
where ${\hat a}_{2L}^e$ is the left-handed coupling between the
$Z_2$-boson and the SM electron (in -$e$ units) and $M_2$ is the
$Z_2$-mass apart from weak corrections (see
\cite{Accomando:2008jh}). We consider $M_2\leq 2$ TeV in order to
agree with the strongest partial wave unitarity bound shown in Fig.
\ref{bi1}. In the dashed region, where only the $Z_2$-resonance  is
visible, the four site Higgsless model could  be  misidentified. A
useful observable to distinguish it from other theories can be the
forward-backward charge asymmetry $A_{FB}$ \cite{Accomando:2008jh}.

The main information one gets from Fig.~\ref{fig:visibilita} is that
the four site Higgsless model can be explored at the LHC in the
favoured Drell-Yan channel, over a large portion of the parameter
space. The same analysis has been performed in the charged channel
\cite{Accomando:2008jh} where the statistical significance can be
about a factor two bigger than in the  neutral one, however the
charged Drell-Yan observables are not that clean. In order to have a
well defined information on the four site Higgsless model
predictions, neutral and charged Drell-Yan channels are thus
complementary. And, more important, there are regions in the
parameter space where they could be both investigated for the search
of all six extra gauge bosons, $W_{1,2}^\pm$ and $Z_{1,2}$, at the
LHC start-up with a luminosity of the order of 1-2 fb$^{-1}$ for
$M_{1,2}\le$ 1 TeV. These results do not include the detector
simulation. At present, this work is in progress, but we do not
expect a drastic change in our conclusions owing to the very clean
signals.

\end{document}